# Running C++ model under Swarm Environment

Richard Leow [2] and Russell K. Standish [1][2]

[1] School of Mathematics
[2] High Performance Computing Support Unit,
University of New South Wales,
Sydney, NSW 2052, Australia.
richardl@hpc.unsw.edu.au    R.Standish@unsw.edu.au
http://www.hpc.unsw.edu.au

*Abstract*

Objective-C [1,2,3,4,5] is still the language of choice if users want to run their simulation efficiently under the Swarm [1] environment since the Swarm environment itself was written in Objective-C. The language is a fast, object-oriented and easy to learn. However, the language is less well known than, less expressive than, and lacks support for many important features of C++ (eg. OpenMP for high performance computing application).

In this paper, we present a methodology and software tools that we have developed for auto generating an Objective-C object template (and all the necessary interfacing functions) from a given C++ model, utilising the Classdesc's object description technology [6,7], so that the C++ model can both be run and accessed under the Objective-C and C++ environments. We also present a methodology for modifying an existing Swarm application to make part of the model (eg. the heatbug's *step* method) run under the C++ environment.

## 1.0    Introduction

Swarm [1] is one of the better-known agent based modelling systems. Swarm consists of a simulation engine built in Objective-C that takes a set of objects called a *Swarm*, and a schedule of actions to perform on them. Swarm defines interfaces that the objects need to adhere to (called *protocols* in Objective-C) in order for the simulation engine to interface to the Swarms. Finally, it provides a suite of visualisation tools or instruments that can be attached to running swarms to observe their behaviours. Swarm programs are hierarchical and the codes are object-oriented. First beta version of Swarm was released in 1995 and was written in Objective-C. In 1999, a Java layer was introduced (release 2.0 onwards). A *COM* interface was proposed [10] which would allow objects written in arbitrary OO languages to be interfaced to swarm, but partially implemented before funding ran out.

The Objective-C syntax is a superset of ANSI Standard-C syntax, and its compiler works for both C and Objective-C source code. C++ was primarily designed as a "better C", incorporated object-oriented support as well as many other improvements over C. C++ is the preferred development language by many because the language is fast, widely deployed, well known, provides rich expressive features and supports generic programming and in-lining. Additionally, the language has an OpenMP support built-in and it can easily integrate with C/FORTRAN. Due to its market presence, vendors have provided highly optimising versions of C++ compilers for their various platforms. Therefore, it is highly desirable to be able to use the C++ (let users code their models in C++) and Objective-C (able to utilise various Swarm's tools) languages in combination to take advantage of what is best in both environments.

In this paper, we present a methodology and software tools that we have developed (utilising the Classdesc class descriptor technology) that enables users to auto generate an equivalent Objective-C object template and all the necessary interfacing functions from a given C++ model. These will enable users to run and access their C++ models under both the C++ and Objective-C environments. We also present a methodology for modifying an existing Swarm application (eg. Heatbugs) to make part of it run under the C++ environment. The key concept is to use a modified Classdesc [6] to parse the user C++ model and to generate a relevant class description file. We then use this file to construct an Objective-C translator to auto-generate an equivalent Objective-C object template and all the necessary interface functions. Our C++→Objective-C interfaces are very efficient as a lot of work is done at compile time by the Classdesc and our approach is much more simpler as compared to Daniels *COM* approach [10] as we only target the C++ and Objective-C environments.

## 2.0     C++ model → Objective-C design strategies

### 2.1     Objective-C basics

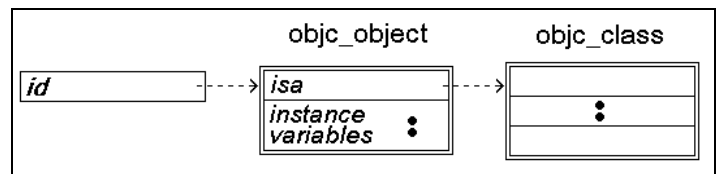

Figure 1   Objective-C object structure

In Objective-C, objects are identified by a distinct data type **id** which is a pointer to the object. Every object carries an **isa** instance variable which points to the object's *class* – what kind of object it is. The *objc_class* structure stores the object's type description. Most objects are derived from the root class object called **Object** - it makes objects behave as Objective-C objects and enable them to cooperate with the run-time system. A *message* is sent to an object to get it to perform useful work (to apply a method). In Objective-C, *message expressions* are enclosed in square brackets (see Figure 2). The *receiver* is an object, and the *message* tells it what to do. In source code, the *message* is simply the name of a method and any arguments that are passed to it.

In Objective-C, messages are not bound to method implementations until run-time. The message function does everything necessary for dynamic binding. It first finds the procedure from the given message of particular receiver, then calls its procedure, passing it the receiving object and the arguments, and finally, returns a value.

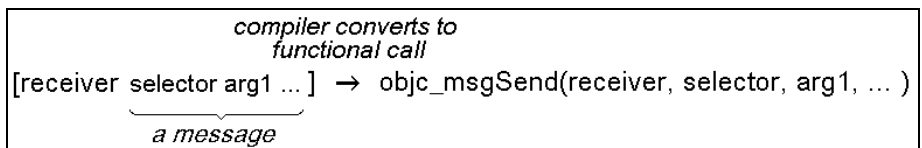

Figure 2   Typical Objective-C message

### 2.2     File conventions used

For an object to be used in both the C++ and the Objective-C environment:
- The *.h* file extension is reserved for the C++ header file and *.mh* file extension is used for the Objective-C header file.
- The *.o* file extension is reserved for the compiled C++ object file and the *.mo* file is used for the compiled Objective-C object file.
- The *.cd* file is the Classdesc output Class Description file from the given object *.h* file. This file is used for the compilation of Objective-C translator (see Section 2.6).

| File name & extension | Purpose | Remarks |
|---|---|---|
| `myobj.h` | `C++ header file` | `User model` |
| `myobj.cc` | `C++ object implementation file` | `User model` |
| `myobj.mh` | `ObjC object header` | `Interface file 1` |
| `myobj.m` | `ObjC object implementation` | `Interface file 2` |
| `myobjExportCpp.cc` | `C++ interface function for ObjC` | `Interface file 3` |
| `myobj.mo` | `Compiled ObjC object` | `ObjC object file` |
| `myobj.o` | `Compiled C++ object` | `C++ object file` |
| `myobj.cd` | `Classdesc description file` | `Classdesc output` |

Figure 3   File conventions used

2.3   Objects creation in the C++ and Objective-C environments

Figure 4 shows an example of a C++ model and Figure 5 shows the equivalent Objective-C object template. Since Objective-C and C++ programs are both run in the same memory space, only one copy of the object is created and shared between both environments.

In our implementation, shared Objective-C/C++ objects are instantiated in the Objective-C environment (since all the objects in Objective-C environment must be inherited from the Objective-C base class for them to behave as Objective-C objects) and are accessed in the C++ environment using the objects' C++ class templates (the Objective-C objects' pointers are passed to the C++ environment for typecasting).

```
#include <stdarg.h>
#include "ObjCsupport.h"          // ObjC supports
#include "vector"

class myCounter: public objc_obj {  // need to be derived from ObjC class
   public:
      char    sName[20];
      int     iaX[2][4];
      . . .

   public:
      int    sum3_x1(int x1, double x2, double x3); // std parameter passing
      double sumN_x1(double x1, objc_t& buf);       // using va_list
      int prtVec(vector<int> v);   // c++ only method – no ObjC translation
      int cpp_prtVec();            // C++ only method due to "cpp_" sub-string
};
```

Figure 4   An example of a user C++ model (*myCounter.h*)

```
#import <objc/Object.h>

@interface myCounter : Object
{  @public
      char sName[20];
      int iaX[2][4];
      . . .
}
- init;
- (double) sum3_x1: (double) x1 x2: (double) x2 x3: (double) x3;
- (double) sumN_x1: (double) x1, ...;
@end
```

Figure 5   An equivalent Objective-C object template (*myCounter.mh*)

## 2.4 Parameter passing between C++ and Objective-C methods

Two methods of parameter passing are supported: the standard method and parameter passing using the Standard-C *stdarg* mechanism [9]. The second method supports an arbitrary number of arguments to be passed to C++ method. Figure 6 shows the two supported parameter-passing methods and examples. Figure 7 shows the implementation details of the required interfacing functions to support these two types of parameter passing.

To aid users in extracting arbitrary number of arguments from the *va_list* structure (through the use of Standard-C **va_start**, **va_arg and va_end** macros), a supporting C++ class called **objc_t** and its corresponding ">>" overloaded output stream operator have been implemented and stored in the in the supporting file *ObjCsupport.h*. An example of the usage is shown in Figure 6 in the user C++ implementation of the *sumN_x1* member method (in this case, to sum a list of three real numbers).

```
Standard parameter passing method:
C++ method :     double sum2_x1(double x1, double x2);
Objective-C:     - (double) sum3_x1: (double) x1 x2: (double) x2;

Parameter passing using va_list:
C++ method :     double sumN_x1(double x1, objc_t& buf);
Objective-C:     - (double) sumN_x1: (double) x1, ...;

Example of Objective-C calls in user main program (main.m):
    double dx = [myObj sumN_x1: 1.1, 2.2, 3.3];

Example of user method implementation in C++ (myCounter.cc):
    double sumN_x1(double x1, objc_t& buf)
    { double x2, x3 ; buf >> x2 >> x3, return x1 + x2 + x3; }
```
Figure 6   Supported parameter passing methods and examples.

```
Objective-C interface functions generated (stored in .m file) :
double cpp_myCounter_sum2_x1(myCounter * obj, double x1 , double x2);
- (double) sum2_x1: (double) x1 x2: (double) x2
{ return cpp_myCounter_sum2_x1(self, x1, x2); }

double cpp_myCounter_sumN_x1(myCounter * obj, double x1, va_list * ap);
- (double) sumN_x1: (double) x1, ...
{ double rtnvalue;  va_list ap;  va_start(ap, x1);
  rtnvalue = cpp_myCounter_sumN_x1(self, x1, &ap);
  va_end(ap);
  return rtnvalue; }

C++ interface functions generated (stored in ExportCpp.cc file):
extern "C"
double cpp_myCounter_sum2_x1(myCounter * obj, double x1 , double x2)
{ return obj->sum2_x1(x1, x2); }

extern "C"
double cpp_myCounter_sumN_x1(myCounter * obj, double x1, va_list * ap)
{ objc_t buffer;  buffer.ap = ap;   return obj->sumN_x1(x1, buffer); }
```
Figure 7   Supporting interfacing functions needed.

Figure 8 shows the algorithm used to translate a C++ function prototype to its Objective-C equivalent. The translation of the C++ function argument string to Objective-C equivalent is defined in the **writeobjc** action in the Classdesc's Objective-C base action definition file *writeobjc_base.h*.

```
Given C++ function prototype :
      Cpp-return-type   function-name   Cpp-argument-list
Output Objective-C equivalent:
      ObjC-return-type  function-name   ObjC-argument-list

construct ObjC-return-type + function name
IF C++-only function
   no-function-translation
ELSE IF (#arg==2 & arg2-type==objc_t&)
   construct va_list-type-ObjC-argument-list
ELSE
   FOR arg₁ to argₙ
      construct standard ObjC-type-argument-list
```

Figure 8    C++ → Objective-C function prototype translation algorithm

To support the use of C++ only member-methods in the user model, following member functions will not be translated to its Objective-C equivalent: argument-list contains any C++ objects (such as vector) or function-name that contains "*cpp_*" sub-string.

2.5    Classdesc modification to parse a user supplied C++ model

Classdesc [6] is a technology that introduces knowledge of an object's structure into the run-time. Classdesc is also part of the *Ecolab* simulation system [7] for studying the evolution of complexity [8]. Both projects are open source projects, and are available for download from the *UNSW HPC* website (http://parallel.hpc.unsw.edu.au). We modified Classdesc to include a *–objc* switch to enable it to output class description definitions that are only relevant to the Objective-C translator. An example of the modified Classdesc output is shown in Figure 9.

```
#include "writeobjc_base.h"
inline void writeobjc(writeobjc_t* targ, eco_string desc,class myCounter& arg)
{
writeobjc(targ,desc+"",(objc_obj&)arg);
writeobjc(targ,desc+".sName",is_array(),arg.sName[0],"[20]");
...
writeobjc(targ,desc+".sum2_x1",arg,&myCounter::sum2_x1, "double", "double x1, double x2");
writeobjc(targ,desc+".sumN_x1",arg,&myCounter::sumN_x1, "double", "double x1 , objc_t & buf");
}
```

Figure 9.   An example of Classdesc output

The output from the Classdesc is stored in the *.cd* file, which consists mostly of **writeobjc** statements (or *actions*). These overloaded **writeobjc** statements will recursive descend the class definition performing the "*translation*" operation on each primitive data type making-up the class. The definitions of writeobjc actions on all primitive data type, including member function type, are defined in the *writeobjc_base.h* file (Classdesc base action definition file).

2.6    The construction of an Objective-C translator

Given the output of a Classdesc class definition file *(.cd* file), the purpose of a translator is to auto-generate an equivalent Objective-C object template and all the necessary Objective-C to C++ interfacing functions. To construct a translator, we need following files: the output file from the Classdesc (eg. *myCounter.cd* file), translator main program file (Figure 10), a user C++ model (eg. *myCounter.h* and *myCounter.cc* files), and the Classdesc base-action definition file (*writeobjc_base.h*).

During the compilation of an Objective-C translator, the class name (or the user model name) is passed to the translator by means of the compiler switch (for example, -DCNAME=myCounter) to enable the translator to include appropriate *.h* and *.cd* files. When the compilation completed, a translator called ***write_objc*** is created. When this translator is executed, it will create an equivalent Objective-C object template and all the necessary interfacing functions.

```
#define WRITE_OBJC(classname) \
writeobjc_headers(&targ, (eco_string) STRING(classname)); \
writeobjc(&targ, (eco_string) STRING(classname), arg);\
writeobjc_tail(&targ)

#include "writeobjc_base.h"
#include FILE_NAME(CNAME,.h)
#include FILE_NAME(CNAME,.cd)
int main()
{  writeobjc_t targ;
   CNAME arg;
   WRITE_OBJC(CNAME);
   return 0;
}
```

Figure 10   Translator main program (*write_objc.cc*)

During the execution of the translator, it creates an instance of the user model (or object) and recursively calls the ***writeobjc*** action on all its member variables and functions for Objective-C translation. When the translator finishes its execution, three interface files are created: the Objective-C object definition file (eg. *myCounter.mh*), the Objective-C object implementation file (eg. *myCounter.m*) and the Objective-C calls to C++ methods interface file (*myCounterExportCpp.cc* file). Since ***writeobjc*** is recursive, it is only necessary to perform this process once, on a top level C++ class.

## 3.0   Four steps for C++ model → Objective-C application generation

Given a C++ model, we need to go through the four steps of transformation (as shown in Figure 11) in order to generate a final application that will run under both the C++ and Objective-C environments. Figure 12 shows the details of the steps involved for final user application generation. Since this can be managed automatically via the *make* process, this complexity is not visible to the programmer.

```
Step 1:        Parse C++ model using Classdesc
Step 2:        Create a translator
Step 3:        Auto-generate interface files
Step 4:        Generate final application
```

Figure 11   Four step for C++ → Objective-C application generation

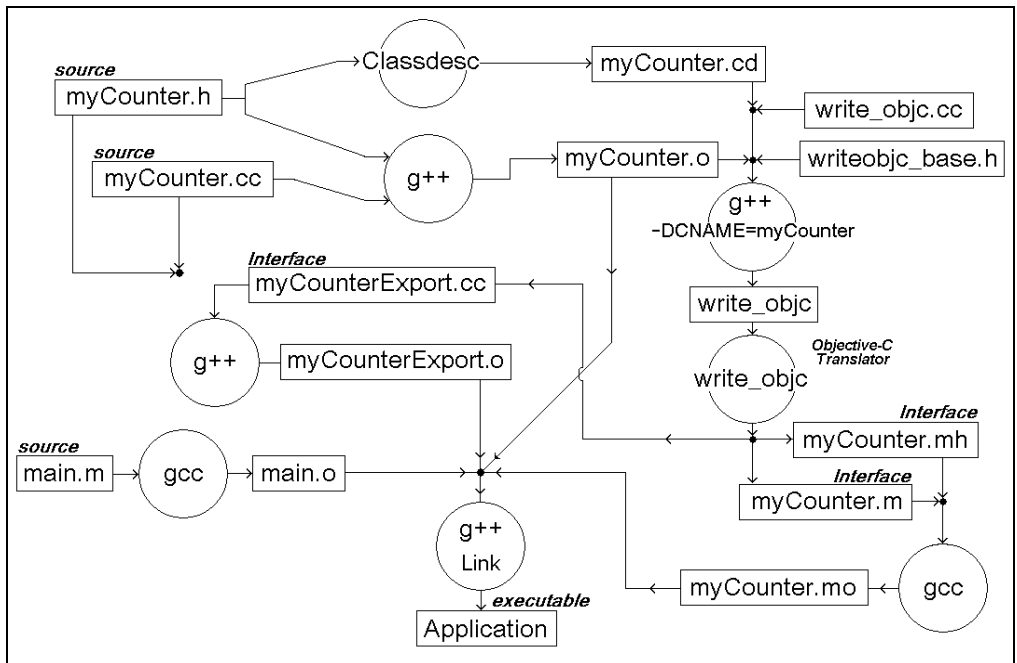

Figure 12 C++ → Objective-C application generation flow diagram

### 3.1 Modification to the Swarm run-time environment

The Swarm run-time environment (see Figure 13) needs to be changed (stored in ~/$(SWARMHOME)/etc/swarm/ sub-directory) to support both the C++ and Objective-C environments.

```
Specify a C++ compiler to use in ~/$(SWARMHOME)/etc/swarm/Makefile.common file
    CPP=/usr/local/gcc2/bin/g++

Add .m.mo and .cc.o rules in ~/$(SWARMHOME)/etc/swarm/Makefile.rule file:
    .SUFFIXES: .o .m .mo .mh .cd .cc .xm
    .m.mo:  $(OBJC) -c -o $@ $(OBJCFLAGS) $(CPPFLAGS) $(DLLCPPFLAGS) $(EXTRACPPFLAGS)
            $(SWARMINCLUDES) $<
    .cc.o:  $(CPP) -g -c $(OPTFLAGS)  $<

Specify the use of C++ for linking in Makefile.appl
    $(APPEXE): $(OBJECTS)
        $(SHELL) $(bindir)/libtool-swarm --mode link $(CPP) $(CFLAGS) $(LDFLAGS) -
```

Figure 13 Swarm environment modification

### 3.2 C++ → Objective-C execution results: myCounter example

Figure 14 shows the Makefile of the *myCounter* example and Figure 15 shows the execution trace of this Makefile and the execution results of the *myCounter* example.

```
.h.cd: classdesc -objc writeobjc < $< > $@
.h.mh: $(CPP) -g -c myCounter.cc
       $(CPP) -g -DCNAME=myCounter -o write_objc $(CPPOBJ) $(OBJC_TRANSLATOR)
       write_objc
.m.mo: $(CC) -c -o $@ -Wno-import $(CFLAGS) $<
.m.o:  $(CC) -c -Wno-import $(CFLAGS) $<
.cc.o: $(CPP) -g -c $(OPTFLAGS)   $<
LIBS = -L/usr/local/lib/gcc-lib/i686-pc-linux-gnu/2.95/ \
       -L/usr/local/lib/libstdc++.so -lgcc -lobjc

OBJC_cd=myCounter.cd
```

```
OBJC_mh=myCounter.mh
CPPOBJ=myCounter.o
INTERFACE_OBJ= myCounterExportCpp.o
OBJC_TRANSLATOR=write_objc.cc
OBJCOBJ=main.o myCounter.mo
OBJ=$(OBJCOBJ) $(CPPOBJ) $(INTERFACE_OBJ)
appl: $(OBJC_cd) $(OBJC_mh) $(OBJ)
      $(CPP) $(CFLAGS) -o main $(OBJ) $(LIBS)

main.o: main.m
myCounter.cd: myCounter.h
myCounter.mo: myCounter.mh myCounter.m
myCounter.o: myCounter.h myCounter.cc
myCounterExportCpp.o: myCounter.h myCounterExportCpp.cc
myCounter.mh: myCounter.h myCounter.cc

clean:
    rm -f *.o *.mo *.*~ *~ main *.cd *,D
    rm myCounterExportCpp.cc write_objc myCounter.mh myCounter.m
...
```

Figure 14   Makefile for myCounter example

```
$ make clean
$ make                                                            ← Step 1
classdesc -objc writeobjc < myCounter.h > myCounter.cd
/usr/local/gcc2/bin/g++ -g -c myCounter.cc                        ← Step 2
/usr/local/gcc2/bin/g++ -g -DCNAME=myCounter -o write_objc myCounter.o
write_objc.cc
write_objc

C++ to ObjC parsing ...                                           ← Step 3
Starts parsing C++ class to ObjC ...
Translating array: "sName[20]" - Translating simple data type: "sName" of-type
"char"
. . .
Translating function: double  sumN_x1(double x1 , objc_t & buf )
. . .
End translation.

gcc -c -Wno-import -g main.m                                      ← Step 4
gcc -c -o myCounter.mo -Wno-import -g myCounter.m
g++ -g -c myCounterExportCpp.cc
g++ -g -o main main.o myCounter.mo myCounter.o myCounterExportCpp.o
    -L/usr/local/lib/gcc-lib/i686-pc-linux-gnu/2.95/
    -L/usr/local/lib/libstdc++.so -lgcc -lobjc

$ main                                                            ← myCounter example
C++ -> ObjC interface testing:
MyCounter C++ Object: byte=<68> ObjName=<c1> dVal=<5.5> iInc=<3>
                     iaX[2][4]=<10 11 12 13 20 21 22 23 >
MyCounter C++ Object: byte=<68> ObjName=<c1> dVal=<8.5> iInc=<3>
                     iaX[2][4]=<10 11 12 13 20 21 22 23 >
Parameter passing using va_list:    Sum = 16.500000
Normal ObjC parameter passing:      Sum = 16.500000
End testing.
$
```

Figure 15   *myCounter* sample execution trace

## 4.0 Objective-C → C++ approach: Heatbugs example

The goal is to make the agent-part of the Heatbugs (the Heatbug's *step* method) run under C++ and leaving the rest to run under the Swarm environment. Currently, Classdesc will not parse the Objective-C code; therefore, we need to manually create an equivalent C++ class template for the Heatbug object (see Figure 16).

```
      Heatbug Objective-C definition       |        Heatbug C++ class template
@interface Heatbug: SwarmObject             class Heatbug
{ @public                                   { public: public objc_obj
                                                 unsigned    zbits;   // fr Swarmobj
    double      unhappiness;                     double      unhappiness;
    int         x, y;                            int         x, y;
    HeatValue   idealTemperature;                HeatValue   idealTemperature;
    HeatValue   outputHeat;                      HeatValue   outputHeat;
    float       randomMoveProbability;           float       randomMoveProbability;
    id <Grid2d> world;                           id          world;
    int         worldXSize, worldYSize;          int         worldXSize, worldYSize;
    HeatSpace   *heat;                           id          heat;
    Color       bugColor;                        Color       bugColor;
}
- step;                                       public:
. . .                                            void step();
@end                                        };
```
Figure 16  Heatbug objective-C definition and its equivalent C++ template

Our approach is to use the techniques that we have developed in previous section for the modification. Firstly, we need to create a C++ Heatbug class template and its implementation file (*Heatbug.h* and *Heatbug.cc*). The C++ Heatbug class will contain one member method called *step* (which contain most of modified Objective-C Heatbug step source code). Also noted that the Objective-C Heatbug object definition uses four user define types: *HeatValue, Color, HeatExtremeType,* and *maxHeat*. Therefore, we also need to type-define them in our C++ class template (Figure 17). The other needed classes are *id* and *objc_obj*, which have already been defined in the *ObjCsupport.h* file. The additional member variable that we needed to include in the Heatbug C++ class template is *zbits*. This *zbits* member variable is used by the Swarm environment for memory zone allocation (the Objective-C Heatbug object inherited it from the **SwarmObject** object and the **SwarmObject** inherited it form the **Object_s** object).

```
typedef int              HeatValue;
typedef unsigned char    Color;
typedef enum {cold,hot}  HeatExtremeType;
extern  const HeatValue  maxHeat;
```
Figure 17  Heatbug related user define type for C++

Next, the C++ Heatbug class needs to export the C++ *step* method for the Objective-C *step* method to call. The exported C++ *step* method is stored in *HeatbugExportCpp.cc* interface file as shown in Figure 18.

Since the Objective-C Heatbug *step* method uses additional Swarm objects (a total of 11 member methods from four different Swarm objects as shown in Figure 19). We also need to export these Objective-C methods to C++ environment (these Objective-C exported functions are stored in *HeatbugExportObjC.m* interfacing file). An example of a C++ calls to the Swarm's member method is shown in Figure 20 (in this case, a C++ *objc_getHeat* interface function calls to the *getHeat* member-method of Swarm's *HeatSpace* object).

```
Objective-C step method (Heatbug.m) :
   void cpp_Heatbug_step(Heatbug * obj);
   - step
   { cpp_Heatbug_step(self);  return self; }

C++ export function (HeatbugExportCpp.cc) :
   extern "C" void cpp_Heatbug_step(Heatbug * obj)
   { obj->step(); }
```
Figure 18   Heatbug Objective-C *step* method modification and interfacing function.

```
Swarm Object       Calls C++ calls Swarm methods (7 methods)
-----------------  ----- ----------------------------------------
SwarmObject              Heatbug object derived from this object
Raster                   GUI part of Heatbug - not modified
Pixmap                   GUI part of Heatbug - not modified

UniformIntegerDist   3   [uniformIntRand getIntWithMin: withMax:]
UniformDoubleDist    1   [uniformDblRand getIntWithMin: withMax:]
Grid2d               1   [world getObjectAtX: Y:]
                     2   [world putObject: atX: Y:]
Diffuse2d            1   [heat getValueAtX: Y:]
                     1   [heat findExtremeType: X: Y:]
                     2   [heat addHeat: X: Y:]
```
Figure 19   Heatbug *step* method uses four swarm objects that needed to be translated to C++.

```
Original Objective-C call:
   int heatHere = [heat getValueAtX: x Y: y]

C++ modification:
   int heatHere = objc_getHeat(heat,x,y);

Example of Objective-C export function (HeatbugExportObjC.m):
   extern int objc_getHeat(void * heatobj, int px, int py)
   { return [(HeatSpace *) heatobj getValueAtX: px Y: py]; }
```
Figure 20   C++ calls to Swarm-objec member method interfacing.

Finally, we need to modify the *Heatbugs* Makefile (Figure 21) to make it compiles both the C++ and Objective-C source codes. Figure 22 shows the execution trace of this Makefile and the execution results of the modified *Heatbugs* sample.

```
...
CPPOBJ= Heatbug.o
OBJCOBJ=Heatbug.mo HeatSpace.o main.o \
    HeatbugModelSwarm.o HeatbugObserverSwarm.o \
    HeatbugBatchSwarm.o HeatbugExportObjc.o
OBJECTS=$(OBJCOBJ) $(CPPOBJ)
...
main.o: main.m Heatbug.mh HeatSpace.h \
    HeatbugObserverSwarm.h HeatbugBatchSwarm.h
Heatbug.mo: Heatbug.m Heatbug.mh
Heatbug.o: Heatbug.cc Heatbug.h
...
HeatbugExportObjc.o: HeatbugExportObjc.m
```
Figure 21   Heatbug Makefile modification.

```
$ make                    ← Heatbugs Makfile execution trace
-c -o Heatbug.mo -g -O2 -Wall ....
gcc -c -g -O2 -Wall -Wno-import ....
...
g++ -g -c Heatbug.cc
```

```
~swarm-2.1.1/bin/libtool-swarm --mode link ....
g++ -g -O2 -o heatbugs Heatbug.mo ....
...
creating heatbugs
$

$ heatbugs                    ← run both in C++ and Objective-C/Swarm environments
T=1 Hb1 at (60,65) feels cold moves to (60,66)
T=1 Hb2 at (39,40) feels cold moves to (40,40)
T=1 Hb3 at (30,7) feels cold moves to (30,6)
...
$
```

Figure 22  Modified Heatbugs run in both C++ and Objective-C environments.

## 5.0  Summary

In this paper, we have demonstrated the use of Classdesc class description technology to parse a user written C++ model and output a class description definitions (save in a *.cd* file). We then use this *.cd* file to construct an Objective-C translator. When this translator is executed, it will auto-generate an equivalent Objective-C template and all the necessary interface functions so that the user C++ model can both be run under the C++ and Objective-C/Swarm environments. In addition, we also developped a methodology of modifying an existing Swarm application to make part of it run under the C++ environment. This will provide the users the capability of coding their models using C++ language while still able to utilise all the software tools available in the Swarm environment. To facilitate the Swarm and C++ integration, we still need to create a C++ layer (C++ Swarm-object class templates and member methods interfacing) for all the existing Swarm objects.

Classdesc is currently an open source project, and latest version is available from the *UNSW HPC*[1] web site. You can also register as a developer by emailing *R.Standish@unsw.edu.au*. This will allow you to access the code as it is being developed, and submit your own code for changes. The project is managed by Peter Miller's *Aegis*[2] code management system, which allows multiple developers to work independently on the code.

---

[1] http://parallel.hpc.unsw.edu.au
[2] http://www.canb.auug.org.au/~millerp/aegis/aegis.html